\documentclass[letterpaper, 10 pt, conference]{ieeeconf}  

\IEEEoverridecommandlockouts                              

\usepackage{mathtools}

\usepackage{amsmath,amsfonts,amssymb,amsthm}
\usepackage{ulem}
\usepackage{float}
\usepackage{subfigure}
\usepackage{xcolor}

\usepackage[bottom]{footmisc}
\usepackage{booktabs, ltablex, siunitx, threeparttablex}
\newcolumntype{C}{>{\centering\arraybackslash}X}

\keepXColumns

\usepackage[ruled]{algorithm2e}

\DeclareMathOperator*{\argminA}{arg\,min}

\begin{document}

\title{\LARGE \bf Design and assessment of an eco--driving PMP algorithm for optimal deceleration and gear shifting in trucks}

\author{B. Wingelaar$^{1}$, G. R. Gon\c{c}alves da Silva$^{1}$, M. Lazar$^{1}$, Y. Chen$^{2}$ and J. T. B. A. Kessels$^{3}$
\thanks{$^{1}$Department of Electrical Engineering, Eindhoven University of Technology, The Netherlands: 
{\tt\small b.wingelaar@student.tue.nl, g.goncalves.da.silva@tue.nl, m.lazar@tue.nl}}%
\thanks{$^{2}$College of Electrical Engineering and Automation, Fuzhou University, China:
       {\tt\small ytchen\_fzu@163.com}}%
\thanks{$^{3}$DAF Trucks, The Netherlands:\newline
       {\tt\small john.kessels@daftrucks.com}}%
\thanks{This research was funded by the European Union’s Horizon 2020 Research and Innovation Programme, Grant Agreement no. 874972, Project LONGRUN.}
}

\maketitle
\pagestyle{empty}

\begin{abstract}
In this paper, an eco--driving Pontryagin maximum principle (PMP) algorithm is designed for optimal deceleration and gear shifting in trucks based on switching among a finite set of driving modes. The PMP algorithm is implemented and assessed in the IPG TruckMaker traffic simulator as an eco--driving assistance system (EDAS). The developed EDAS strategy reduces fuel consumption with an optimized velocity profile and, in practice, allows contextual feedback incorporation from the driver for safety. Furthermore, the optimization over driving modes is computationally inexpensive, allowing the methodology to be used online, in real--time. Simulation results show that significant fuel savings can be achieved proportional to the number of velocity events and the difference between current velocity and final desired velocity for each event.

\end{abstract}

\section{Introduction}
Trucks 
are responsible for about 27\% of the road transport CO$_2$ emissions in the European Union \cite{euemissionswebsite} and thus must be part of the process of fossil fuels reduction. Besides, an increased fuel saving also provides a financial incentive, both for internal combustion engines (ICE) and electrical or hybrid motors (longer trips before recharging batteries). Fuel savings can be obtained by improving vehicle components, but also by applying eco--driving (ED) strategies. Eco--driving is a driving strategy that aims to minimize fuel consumption and emissions by determining an optimal velocity profile, which can be defined as an optimal control problem (OCP) \cite{sciarretta, ifacpaper, PAREDES2019556}. 

Advanced driver--assistance systems can be employed to provide ED advice to drivers in real--time, online \cite{sciarretta}, which is referred to as eco--driving assistance systems (EDAS). The advice can include optimal velocity profiles, gear position and driving modes, and it can be delivered utilizing speech, visual interfaces and haptics.
EDAS has three main benefits: fuel consumption is typically reduced; the driver decides to implement the advice based on situational contextual feedback \cite{ifacpaper,edassimulator}, which accounts for safety; and it does not require complex powertrain modifications \cite{ifacpaper,complexsource}. Moreover, ED has also been studied for hybrid electric powertrains in \cite{ZHU2019562}, for full electric powertrains in \cite{gao2019evaluation,lee2020model}, and ED has been extended with the information of curvature road sections in \cite{curvedroad} and signalized intersections in \cite{sigint}. Extensions that consider small platoons can be found in \cite{sharma2020optimal, chen2018real}. The non-convexity of the discretized ED--OCP for an electric HDT has been addressed in \cite{padilla2018global}, where a reformulation 
has been proposed such that, if the problem is feasible, then it has a unique global minimum. 


Recent studies \cite{thomassen,nazar,ifacpaper} have shown that eco--driving with optimization over driving modes, rather than low-level continuous control variables (e.g., engine torque, brake torque or preceding vehicle speed) \cite{7795650,SCHORI2013109,powah}, can reduce fuel consumption and solve the OCP in a computationally inexpensive way. 
A simplified hybrid powertrain model (without gears and state of charge (SoC) modeling) is studied in \cite{nazar} for velocity profile optimization using a fast Pontryagin’s Maximum Principle (PMP) algorithm, which can run online in a truck. In \cite{ifacpaper}, a model predictive control (MPC) approach was applied to a detailed ICE powertrain model, including gear shifting as a decision variable, for the same optimal control problem. However, the inherent online computing load of a real--time MPC approach in general can become challenging~\cite{7393561}.

Performance assessment of the results has also been a challenge in the field of ED due to the lack of real--time performance data or the need to construct complex traffic scenarios for detailed simulations. A few studies have implemented ED on actual vehicles and even fewer on heavy-duty trucks \cite{8917077}, and numerical or microscopic traffic simulations can be found in \cite{ifacpaper,nazar,thomassen}. Thus, it is of interest to assess ED algorithms implemented as EDAS in a realistic simulator for trucks, with a sufficient amount of data to provide a reliable performance assessment. 


In this paper we adopt the ED approach in \cite{thomassen,nazar} that optimizes over driving modes and we design a fast PMP algorithm for optimal deceleration and gear shifting in ICE trucks. Compared to existing works, the developed PMP algorithm features the following innovations/improvements: \emph{(i)}--new co--state initialization algorithm with better convergence; \emph{(ii)}--incorporation of ICE constraints violation for obtaining feasible solutions and \emph{(iii)}--inclusion of gear shifting as decision variables. Also, we develop more realistic supporting models for each driving mode, e.g., including  powertrain rotational inertia, as this can represent high additional load \cite{homberg,saerens}. The corresponding EDAS is implemented in IPG TruckMaker \cite{TM} and tested \emph{online} on the European transient cycle (ETC) driving cycle \cite{etcbook}. 

\section{Driving modes modeling and PMP algorithm}
\label{sec:methods}


This section presents the longitudinal dynamics model and the formulation of the driving modes. Then, the ED optimal control problem (OCP) is formulated and the PMP algorithm is introduced to solve this problem.

\subsection{Longitudinal dynamics}\label{sec:longdyn}
For the scope of this paper, we only consider the longitudinal dynamics of a truck, consisting of a simplified powertrain with ICE, clutch, gearbox, final drive, and rear wheels. Furthermore, an ideal clutch is assumed with no slip and a 100\% transmission efficiency. We introduce the following driving modes: cruising, eco-roll, coasting and engine brake.  Unlike other studies, no universal longitudinal dynamics equation is given that is shared between all the driving modes. Instead, an individual dynamics equation is presented for each driving mode. 
The basis for the fundamental dynamics can also be found in \cite{ifacpaper,sciarretta,nazar,thomassen} and extensions that also include rotational inertia in \cite{homberg,saerens}.

The dynamics are based on Newton's second law of motion and formulated in the spacial distance domain $s\in\mathbb{S}$, instead of the time  domain. This complies with earlier studies \cite{ifacpaper, nazar,thomassen}, as relevant boundary conditions such as road grade, speed limits and traffic (signals) are defined in this domain. The dynamics conversion from time domain to spacial domain is defined as follows:
\begin{equation}
        \frac{dv}{ds}=\frac{dv}{dt}\frac{dt}{ds}=\frac{\sum F}{m}\frac{1}{v},\\
\end{equation}
where $v$ is the state variable (velocity), $m$ the equivalent mass of the truck and $\sum F$ the net force. 

Subsequently, we define elements which the driving modes have in common. 
The engine speed $\omega_e$~[RPM] depends on the engagement of the ICE with the drivetrain and is defined as: 
  \begin{equation}
    \omega_e=
    \begin{cases}
      \frac{30 i_t(y)i_rv}{\pi r_w} & \text{if the clutch engaged}\\
      \omega_{idle} & \text{otherwise},
    \end{cases}
  \end{equation}
where $i_t(y)$ is the gear ratio as function of gear $y$, $i_r$ the final drive ratio, $r_w$ the radius of the rear wheels and $\omega_{idle}$ is a constant idle speed to keep the engine running. The total resistance force $F_{res}$ is also considered in all the driving modes and is modeled as:
\begin{equation}
      F_{res} = \tfrac{1}{2}\rho_ac_dA_fv^2 + mgc_r\,\text{cos}(\alpha) + mg\,\text{sin}(\alpha),
\end{equation}
where $\rho_a$ is the air density, $c_d$ the aerodynamic drag coefficient, $A_f$ the frontal area of the truck, $g$ the gravitational constant, and $c_r$ the roll resistance coefficient. The parameter $\alpha$ represents the road slope in [rad] and it is assumed that $\alpha\approx0$, thus road gradients are negligible. 
The values for these parameters correspond to predefined vehicle parameters available within the IPG TruckMaker software \cite{TM} and they are provide in the Appendix.

Conversely to the drivetrain model in \cite{ifacpaper} where $T_e$ represents the indicated torque, we consider $T_e$ as the engine output torque that is measured by a dynamometer, which already includes the ICE internal friction torque $T_{int.fric}$. Additionally, we include the rotational parts equivalent inertia of the truck in the model. 


\subsubsection{\textbf{Cruising mode $M^{cr}$}}
\begin{figure}[htb]
    \centering
    \includegraphics[width=0.95\linewidth]{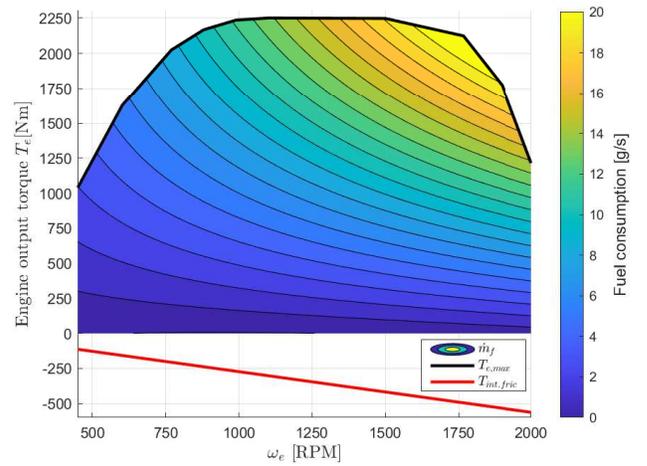}
    \caption{Fuel consumption map  $\dot{m}_f$~[g/s], the maximum engine torque $T_{e,max}$~[Nm] and internal friction torque $T_{int.fric}$~[Nm] of the ICE.}
    \label{fig:fmap}
\end{figure}

In cruising mode, the truck drives with a constant velocity in gear position $y>0$ and the clutch is engaged with the drivetrain. The dynamics in this mode are described by:
\begin{equation}
   \frac{\text{d}v}{\text{d}s}= \frac{1}{mv}\left(\frac{i_r i_t(y)}  {r_w}T_e-F_{res}\right)=0.
\end{equation}
The corresponding fuel consumption is given by:
\begin{equation}
   \frac{\text{d}m_f}{\text{d}s}= \frac{1}{v}\dot{m}_f(T_e,\omega_e),
\end{equation}
where $\dot{m}_f$ is a polynomial fit of the fuel consumption in [g/s] as function of $T_e$ and $\omega_e$ that is formulated as:
\begin{equation}
    \dot{m}_f = \beta_0+\beta_1\omega_e+\beta_2 T_e+ \beta_3\omega_e^2+\beta_4\omega_e T_e+\beta_5 T_e^2,
\end{equation}
where $\beta_0$,...,$\beta_5$ are the polynomial coefficients. Fig. \ref{fig:fmap} illustrates this polynomial as well as the ICE constraints and  the ICE internal friction torque $T_{int.fric}=(\gamma_0+\gamma_1 \omega_e)$.

\subsubsection{\textbf{Eco--roll mode $M^{ec}$}}
In  the eco--roll mode the clutch is disengaged from the rest of the drivetrain ($T_e=0$) with the gear position in neutral ($y=0$). Resistance forces decelerate the truck, while the equivalent rotational inertia of the drivetrain reduces deceleration. The dynamics are described by:
\begin{equation}
    \frac{dv}{ds}=-\frac{r_w^2}{(mr_w^2+J_{eq.dt})v} F_{res},
\end{equation}
where $J_{eq.dt}$ is the equivalent rotational inertia of the drivetrain in neutral gear. The disengaged ICE rotates at an idling speed $\omega_{idle}$ and consumes a small, constant amount of fuel:
\begin{equation}
    \frac{\text{d}m_f}{\text{d}s}=\frac{1}{v}\dot{m}_{f,idle}.
\end{equation}

\subsubsection{\textbf{Coasting mode $M^{co}$}}
In coasting mode, because the clutch is engaged, the resistance force as well as the internal engine friction decelerate the truck, but similarly to the eco--roll mode, the equivalent inertia reduces deceleration. The dynamics are described by:
\begin{equation}
        \frac{dv}{ds}=-\frac{r_w^2}{(mr_w^2+J_{eq.pt}(y))v} \left(\frac{i_r i_t(y)}  {r_w}T_{int.res}+F_{res}\right),
\end{equation}
where $J_{eq.pt}(y)$ is the combined equivalent rotational inertia of the wheels and rotating parts of the powertrain, dependent on gear position $y$. No fuel is consumed in this mode.

\subsubsection{\textbf{Engine brake mode $M^{eb}$}}
The ICE can deliver an additional braking force with a continuous service brake. We assume an ideal continuous service brake that does not consume any energy nor fuel when enabled in this driving mode. The dynamics are described by:
    \begin{equation}
        \frac{dv}{ds}\!=\!-\frac{r_w^2}{(mr_w^2\!+\!J_{eq.pt}(y))v}\!\left(\!\frac{i_r i_t(y)}  {r_w}(T_{int.res}\!+\!T_{eb})\!+\!F_{res}\!\right)\!,
\end{equation}
where $T_{eb}$ is the braking torque of the service brake.

\subsection{OCP formulation and PMP algorithm}\label{sec:PMP}




%
%
%
Switching between the driving modes requires switching between different continuous dynamics. Thus, the corresponding OCP contains both discrete and continuous variables, which can be classified as a hybrid optimal control problem \cite{SCHORI2013109}, which can be solved by applying an extension of the PMP to hybrid problems \cite{DMITRUK2008964}. The OCP with constraints is defined as:
\begin{subequations}\label{eq:OCP}
\begin{align}
  &J=\min_{M^i\in\tilde{M}, i=0,\ldots,L }\;\sum_{i=0}^{L}\int_{s_i}^{s_{i+1}}g(M^i)\text{d}s\label{eq:costHOCP}\\
  &g(M^i)=w_f\frac{1}{v}\dot{m}_f(M^i)+w_t\frac{1}{v}\label{eq:gM}\\
  &\tilde{M}=\{M^{cr,y},M^{ec},M^{co,y},M^{eb,y}\}\label{eq:Mset}\\
  & s.t. ~~v(s_0)=v_{0}, ~~~~ v(s_f)=v_{f}\label{eq:c1}\\
  & ~~~~~~\omega_{e,min}(T_e)\leq\omega_e(y,v)\leq\omega_{e,max}(T_e)\label{eq:c2}\\
  & ~~~~~~ T_e \leq T_{e,max}(\omega_e)\label{eq:c3}.
\end{align}
\end{subequations}

The continuous--time cost function \eqref{eq:costHOCP}--\eqref{eq:gM} is a cost placed on time duration and fuel consumed with tuning parameters $w_t$ and $w_f$ respectively. $L$ represents a finite number of road segments, with $M^i\in\tilde{M}$ the driving mode on each road segment, where \eqref{eq:Mset} is the  set of (sub--) driving modes. In total there are $3n+1$ driving modes for $n>0$ gear positions. Furthermore, the OCP is constrained by the boundary conditions \eqref{eq:c1} and by the ICE operational conditions \eqref{eq:c2} and \eqref{eq:c3}. 
The Hamiltonian and the (co--) state dynamics are given by:
\begin{subequations}
\begin{align}
        H_{M^i}(v,\lambda)&=\lambda^\intercal f_{M^i}(v,\lambda)+g(M_i), \label{eq:conH}\\
            \frac{dv}{ds}&=\frac{\partial H_{M^i}(v,\lambda)}{\partial \lambda},\\
                \frac{d\lambda}{ds}&=-\frac{\partial H_{M^i}(v,\lambda)}{\partial v},
\end{align}
\end{subequations}
where $H_{M^i}$ is the Hamiltonian, $f(M^i)$ denotes the state dynamics $\frac{dv}{ds}$ as a function of the driving mode active on a road segment and $\lambda$ is the co--state. 
The problem is then discretized into $N$ distance samples of length $\delta s$ on which it is assumed that each distance sample has one constant driving mode, i.e.
\begin{equation} \label{eq:distancediscretization}
    J = \sum_{k=0}^{N-1} g_{M_k}(v_k), ~~~~ \delta s=\frac{s_f-s_0}{N}.
\end{equation}
Then, the (co--)state on every distance sample, which is piecewise continuous, is also discretized with the forward Euler method. The discretized Hamiltonian and the backward (co--)state \textit{update} is given by
 \begin{subequations}
\begin{align}
 H_{M_k}(v_k,\lambda_k)&=\lambda_k^\intercal f(v_k)+g_{M_k}(v_k),\label{eq:discreteH}\\
        v_{k-1} &= v_k-\frac{\partial H_{M_k}(v_k,\lambda_k)}{\partial \lambda}\delta s, \label{eq:updatediscretestate}\\
        \lambda_{k-1} &= \lambda_k+\frac{\partial H_{M_k}(v_k,\lambda_k)}{\partial v}\delta s. \label{eq:updatediscretecostate}
\end{align}
\end{subequations}
The optimal driving mode $M_k^*$ on distance sample $k$ then has the minimizing Hamiltonian:
\begin{equation}
    M_k^* = \argminA_{M_k\in\tilde{M}}H_{M_k}(v_{k+1},\lambda_{k+1}).\label{eq:optargmin}
\end{equation}
This optimal driving mode is determined for every distance sample in a backward order for a fixed terminal co--state $\lambda_N:=\lambda_{k=N}$, as summarized in Algorithm~\ref{alg:opt}.
\begin{algorithm}[!h]
    \SetAlgoLined
    \textbf{Input:} $N$, $s_0$, $s_f$, $v_0$, $v_f$, $\lambda_N$ // (all scalars)\\
        \textbf{Output:} $y$, $v$, $M$ // (all vectors) \\
     \For{$k = N,N-1\dots1$}{
     $H=$ calculateHamiltonians($\lambda_k$,$v_k$) // apply \eqref{eq:discreteH}\\
     $H=$ deleteICEconstraintsViolatingEntries($H$) // remove entries violating \eqref{eq:c2} and \eqref{eq:c3}\\
     $[M_{opt},y_{opt}]=\arg\min(H)$ // select Hamiltonian minimizing driving mode and gear using \eqref{eq:optargmin}\\
     $y_k=y_{opt},\quad M_k=M_{opt}$ // store data\\
     $v_{k-1}=$ updateState($v_k$,$\lambda_k$,$M_{opt}$) // apply \eqref{eq:updatediscretestate}\\
     $\lambda_{k-1}=$ updateCostate($v_k$,$\lambda_k$,$M_{opt}$) // apply \eqref{eq:updatediscretecostate}\\
    }
        \textbf{return} $y$, $v$, $M$\\ 
\caption{Discrete PMP optimization algorithm}\label{alg:opt}
\end{algorithm}

Compared with existing PMP algorithms for eco--driving \cite{thomassen,nazar}, which only consider initial and final velocity constraints, before computing \eqref{eq:optargmin}, Algorithm~\ref{alg:opt} eliminates all entries violating constraints \eqref{eq:c2} and \eqref{eq:c3} at current iteration $k$.
Note that the choice of $\lambda_N$ fully determines the state trajectories and initially, the value of $\lambda_N$ is unknown. \cite{nazar} proposes a recursive update of an initial guess for $\lambda_N$ proportional to the difference between the actual current velocity $v_0$ and the calculated initial velocity $v(s_0)$.

Next, we present a new iterative procedure for optimizing the value of $\lambda_N$ within predefined upper and lower bounds, as summarized in Algorithm~\ref{alg:lambdaN2}. Therein, the sign of the error between the $v_0$ and $v(s_0)$ in combination with bisection is exploited to recursively update predefined upper and lower bounds on $\lambda_N$. Running Algorithm~\ref{alg:lambdaN2} extensively in various simulations indicates significant improvement in convergence compared to the update rule derived in \cite{nazar}.
\begin{algorithm}[!h]
    \SetAlgoLined
    \textbf{Input:} $v_0$, $v_f$, $\lambda_{N,min}^0$, $\lambda_{N,max}^0$, maxIter, tol\\
    \textbf{Output:} $y$, $v$, $M$ // (all vectors)\\
    $\lambda_N^0=0$, $\text{error}=\text{tol}+1$; $i=0$;\\
    \While{(i $\leq$ maxIter) and $(|\text{error}|>\text{tol})$}{
$[y, v, M]$ = runPMPopt($\lambda_N^i,v_0,v_f,...$) // Alg. \ref{alg:opt} \\
    e = $v_0-v(s_0)$    // define error\\
    $\lambda_{N,max}^{i+1}$ = sign(e)\,max(sign(e)$\lambda_{N,max}^{i}$, sign(e)$\lambda_{N}$)\\
    $\lambda_{N,min}^{i+1}$ = sign(e)\,max(sign(e)$\lambda_{N,min}^i$, sign(e)$\lambda_{N}$)\\
    $\lambda_{N}^{i+1}=\tfrac{1}{2}(\lambda_{N,min}^{i+1}+\lambda_{N,max}^{i+1})$\\
    $i=i+1$\\
    }
\textbf{return} $v$, $y$, $M$\\ 
\caption{$\lambda_N$-algorithm 
}\label{alg:lambdaN2}
\end{algorithm}
The idea of recursively updating upper and lower bounds on the co--state $\lambda_N$ has also been utilized in \cite{chen2018real} for optimizing a cost function with a terminal penalty term.
\begin{figure}[H]
    \centering
    \includegraphics[width=\linewidth]{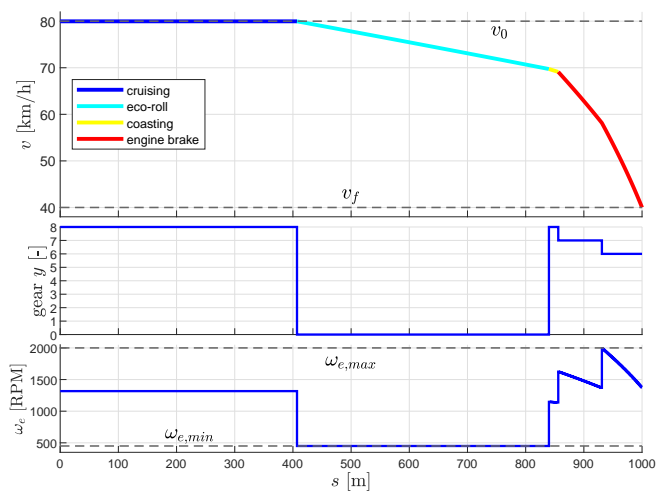}
    \caption{PMP algorithm output: (top) optimized velocity--distance profile consisting  of the driving modes; (middle) corresponding gear position; (bottom) corresponding engine speed.}
    \label{fig:pmpoutput}
\end{figure}
An example of the output of Algorithm~\ref{alg:opt} is illustrated in Fig.~\ref{fig:pmpoutput} as a sequence of optimal driving modes, corresponding velocity profile, gear positions and engine speed over a road segment. It can be observed that the resulting engine speed is always within constraints, which is not guaranteed in general by other ED algorithms based on optimal control, including the PMP algorithm in \cite{nazar}.

\section{Implementation in IPG TruckMaker}\label{sec:tmsec}
\begin{figure}
    \centering
    \subfigure[]{\includegraphics[width=0.48\linewidth,height=0.48\linewidth]{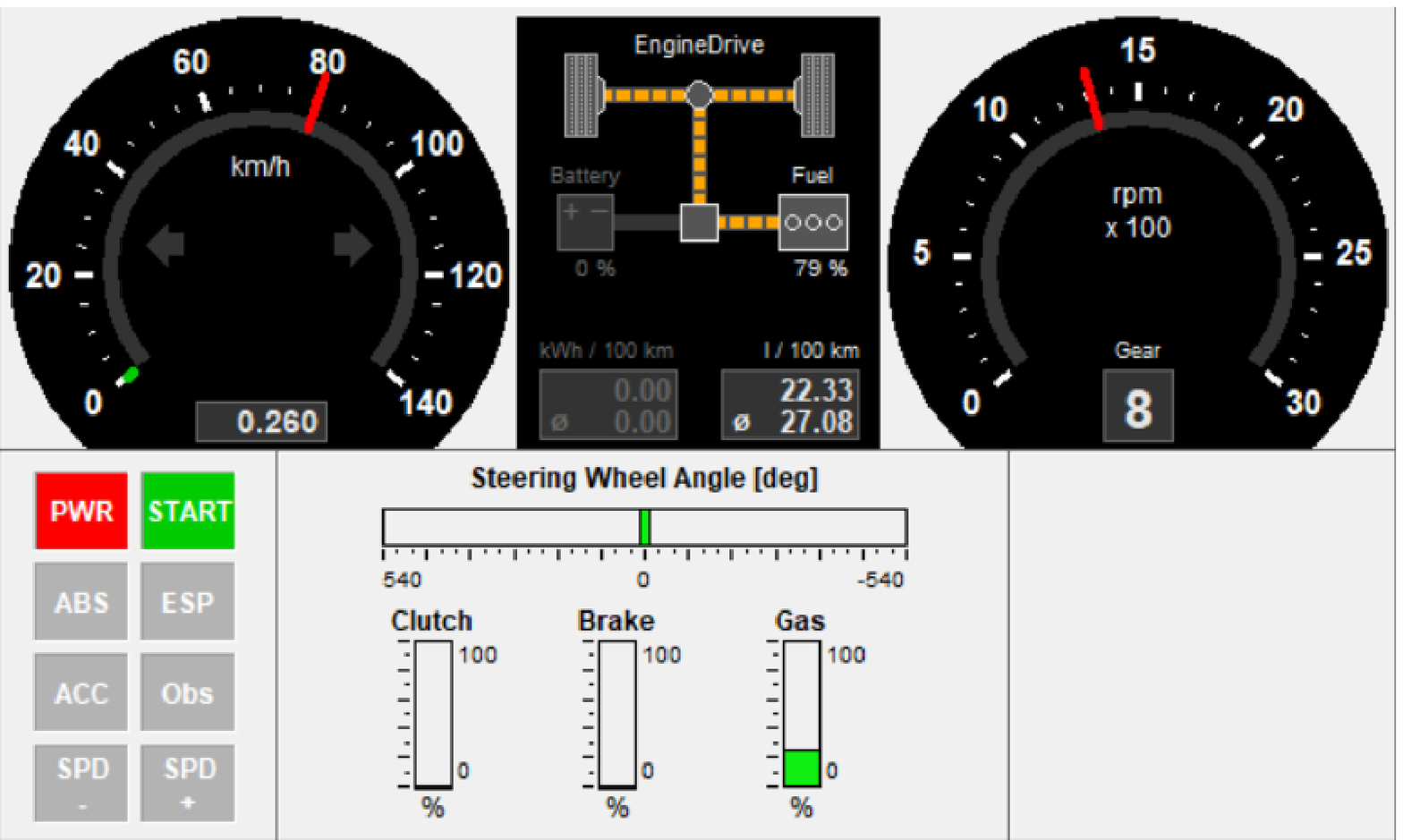}}
    \subfigure[]{\includegraphics[width=0.48\linewidth,height=0.48\linewidth]{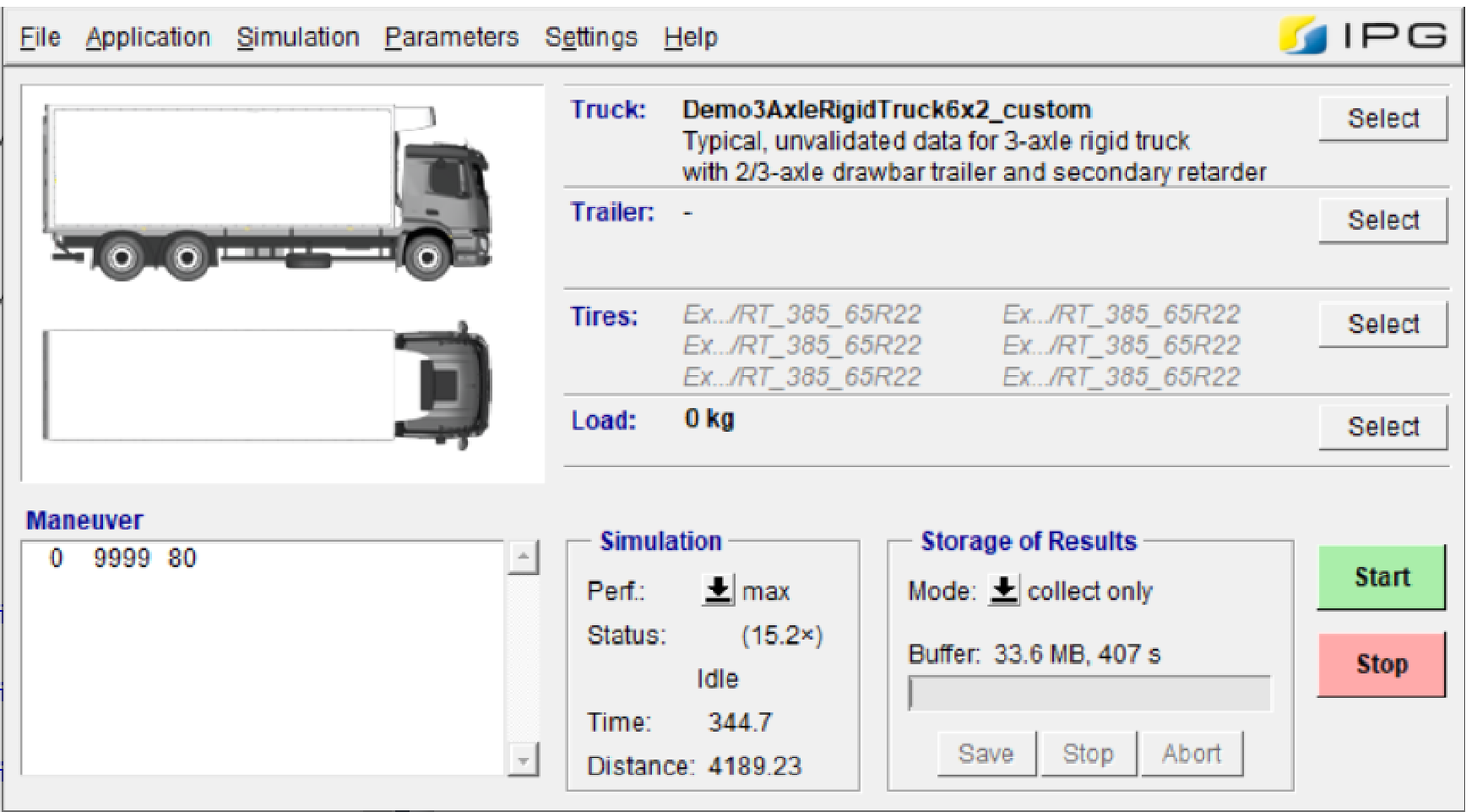}}
    \caption{IPG TruckMaker interface: (a) Instruments panel including pedal positions, speedometer and tachometer; (b) IPG TruckMaker main GUI.}
    \label{fig:TMimpresion}
\end{figure}

In this paper we are using IPG TruckMaker \cite{TM} in combination with Matlab--Simulink for simulations and performance assessment of the developed EDAS. In TruckMaker, one can simulate the vehicle dynamics of rigid and articulated trucks and busses amongst other vehicles, with a range of customization options. For the simulations, traffic scenarios with custom environments can be designed and real--time data can be logged. An instrument panel can also be displayed to provide additional info. Fig.~\ref{fig:TMimpresion} provides an instance of the IPG TruckMaker graphical user interface (GUI).

The IPG TruckMaker simulation setup was designed for the scope of the proposed algorithm. Since only the longitudinal dynamics of rigid body truck in the $s$--domain are considered, a straight road with zero road gradient was designed with a speed limit and a trigger. In the real world, this trigger could be activated by a GPS coordinate, a certain threshold exceeded by the radar sensor of the truck or a signal from an eco--routing planner. Additionally, the primary retarder option was selected as the continuous service brake.

\begin{figure}[htb]
    \centering
    \includegraphics[width=1\linewidth]{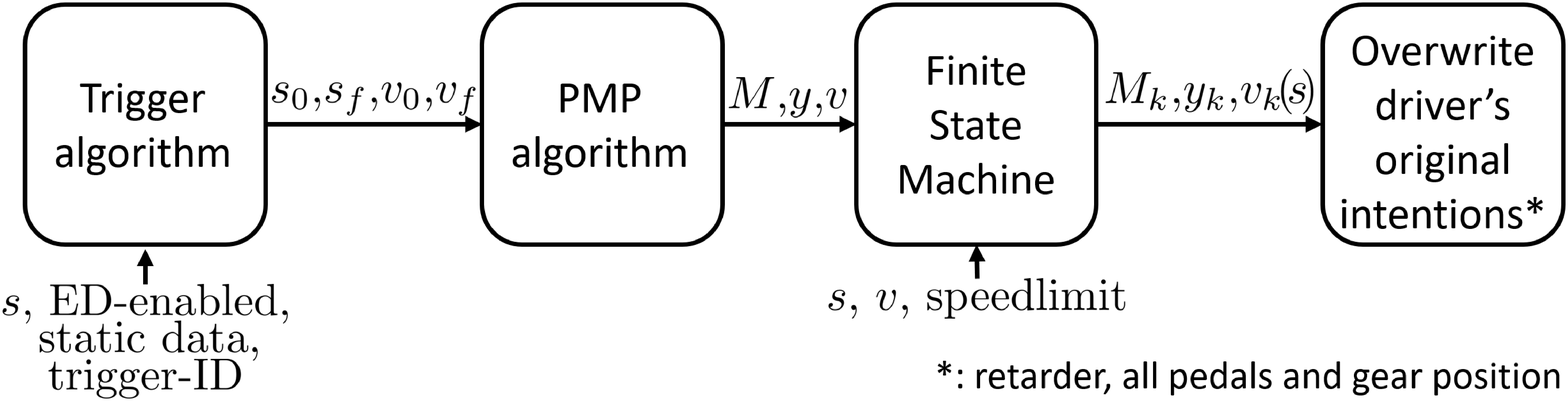}
   \caption{IPG TruckMaker EDAS architecture.}
    \label{fig:overview}
\end{figure}

Fig.~\ref{fig:overview} provides a schematic overview of the EDAS architecture implemented in IPG TruckMaker, which corresponds to Algorithm~\ref{alg:opt} and Algorithm~\ref{alg:lambdaN2}. The eco--driving PMP algorithm is enabled once a velocity trigger event is activated and if the eco--driving function is enabled. The PMP algorithm then computes a sequence of optimal velocities, driving modes and gear positions for a road segment of fixed length. The stored sequence (advice) is transferred to a finite state machine (FSM) which selects and updates the advice for the driver over the entire road segment, based on position and velocity. This advice is then implemented by overwriting the IPG--driver (gear position, retarder enabled and all pedals). If the truck crosses the speed limit or if the final velocity is reached earlier, the FSM eco--driving advice is disabled and the vehicle switches to cruising at the corresponding velocity. Since the vehicle dynamics are deterministic and no other disturbances are present, the PMP algorithm is executed once, when the velocity event is activated. Then the advice is implemented for the complete corresponding road segment. In a real--life traffic scenario, however, the PMP algorithm can be executed in a receding horizon fashion, with a shrinking horizon, as proposed in \cite{nazar}.


\section{Simulation results and assessment}\label{sec:RandD}
In this section, the performance of the PMP algorithm implemented in TruckMaker is evaluated under two test--case scenarios: a highway exit and the ETC~\cite{etcbook}.
\subsection{Test case 1: motorway to urban area}
\begin{figure*}[htb]
    \centering
    \includegraphics[width=1\linewidth]{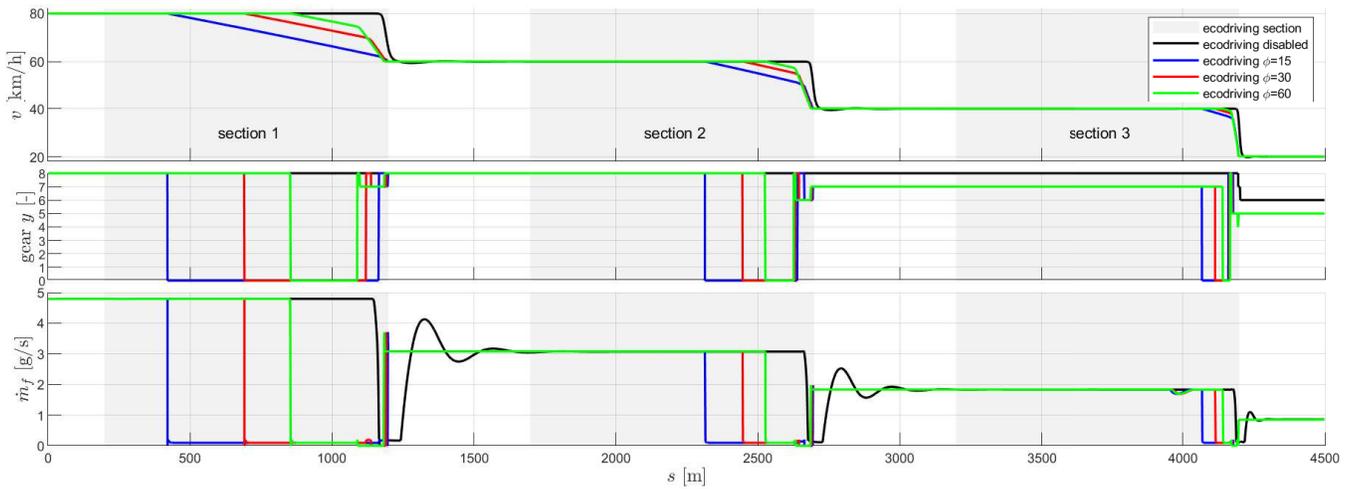} 
    \caption{Test case 1: motorway to urban area route with 3 eco--driving sections with ED disabled and ED enabled with cost weights ratio $\phi$ equal to $15$, $30$ and $60$, respectively: (top) velocity profile, (middle) gear positions, (bottom) fuel consumption.}
    \label{fig:tc1fig}
\end{figure*}

In this test case we use a custom--designed route to assess the performance of the PMP algorithm itself on individual ED sections at different velocities and parameters.
A 4500~[m] long route is considered with 3 deceleration eco--driving sections of 1000~[m] each and with a 20~[km/h] velocity difference on which a driver can be advised. This resembles the route from exiting a motorway to an urban area. 
Further details of the route can be observed in Fig.~\ref{fig:tc1fig}. 
Performance is evaluated for 3 different cost--function weights, such that their ratio $\phi := w_{t}/w_{f}$ is equal to $15$, $30$ and $60$, respectively.
This means that in the optimized velocity--distance profile, the time duration penalty is $\phi$ times greater than the fuel consumption penalty.

Velocity and fuel consumption profiles of a 25200~[kg] truck are plotted in Fig.~\ref{fig:tc1fig}. If eco--driving is disabled, no advice is given to the driver. 
The IPG--driver then reacts by maximizing the allowed velocity and shows anticipating behavior in a certain preview time. A larger difference between the enabled and disabled eco--driving advice behavior can be observed in the fuel consumption profiles between the eco--driving sections where the driver is not advised in either case. Additional nominal and relative data are presented in Table~\ref{tab:T1_full}.
\setlength{\tabcolsep}{4pt}
\begin{table*}[htb]
\caption{Test case 1: trip time increase, fuel consumption and saved fuel for each eco--driving section and the complete route\\ (trip time increase and saved fuel are calculated with respect to their corresponding values for ED disabled).}
\resizebox{\textwidth}{!}{%
\begin{tabular}{cc|cccc|cccc|cccc|cccc}
\hline
         &        & \multicolumn{4}{c|}{section 1} & \multicolumn{4}{c|}{section 2} & \multicolumn{4}{c|}{section 3} & \multicolumn{4}{c}{route}       \\ \cline{3-18} 
ED       & $\phi$ & time  & incr. & fuel   & saved & time  & incr. & fuel   & saved & time  & incr. & fuel   & saved & time   & incr. & fuel   & saved \\
      & {[}-]    & [s]   & [\%]  & [g]    & [\%]  & [s]   & [\%]  & [g]    & [\%]  & [s]   & [\%]  & [g]    & [\%]  & [s]    & [\%]  & [g]    & [\%]  \\ \hline
disabled & [-]    & 45.11 & [-]   & 207.63 & [-]   & 60.11 & [-]   & 179.75 & [-]   & 90.06 & [-]   & 162.83 & [-]   & 332.82 & [-]   & 801.13 & [-]   \\
enabled  & 15     & 49.76 & 10.32 & 52.10  & 74.91 & 62.53 & 4.02  & 116.54 & 35.17 & 91.31 & 1.38  & 144.11 & 11.49 & 341.21 & 2.52  & 576.01 & 28.10 \\
enabled  & 30     & 47.12 & 4.46  & 109.84 & 47.10 & 61.50 & 2.32  & 140.74 & 21.71 & 91.03 & 1.07  & 151.33 & 7.06  & 337.27 & 1.34  & 665.16 & 16.97 \\
enabled  & 60     & 46.38 & 2.82  & 145.25 & 30.04 & 61.24 & 1.87  & 155.38 & 13.56 & 90.95 & 0.98  & 155.53 & 4.48  & 336.18 & 1.01  & 719.42 & 10.20 \\ \hline
\end{tabular}%
}
\label{tab:T1_full}
\end{table*}

Fig.~\ref{fig:tc1fig} and Table~\ref{tab:T1_full} illustrate that a smaller $\phi$ makes the truck decelerate earlier on each ED section, which is accompanied by a lower fuel consumption compared to the case without advice. For a constant $\phi$, the fuel savings increase with a higher velocity and that this comes at an increased cost of extra time duration. Also, the relative fuel savings and time increase on the full route do not represent the averages of the 3 eco--driving sections combined. This can be explained by the fact that most of the time is spent on ED section 3, while the least fuel is saved on that section and the opposite holds for ED section 1. That is, a larger percentage of time and distance must be spent in cruising mode than in section 1.
Thus, the maximum potential for fuel savings is always higher at higher velocities because the fuel consumption per length unit is higher due to increased friction and aerodynamic drag.\par
Observe that in the eco--roll mode, the resistance force is dominated by the roll--resistance force, which is independent of the velocity. In the coasting and the engine brake modes the PMP algorithm minimizes the Hamiltonians by maximizing the engine speed $\omega_e$ (via increased gear ratio). This makes $T_{int.fric}$, which increases with $\omega_e$, approximately constant. Additionally, the engine brake torque of the retarder is also approximately independent of the engine speed. At lower velocities (accompanied by lower gears), although the deceleration effect of these braking torques is reduced by the increased rotational inertia of the powertrain, the increased deceleration effect of the greater gear ratio outweighs the effect of this inertia increase, resulting in an overall deceleration increase. In turn, this results in reduced fuel savings. 

A lower $\phi$ for lower velocities can provide a solution to reduce this inability to make significant fuel savings at low velocities on long distances. For instance, selecting a $\phi$ of 60 on section 1 and a $\phi$ of 15 on section 3 results in more consistent fuel savings with a more gradual time increase. 
In the process of varying $\phi$, it was observed that very low values of $\phi$ may result in an infeasible solution to the OCP problem, while a feasible solution exists. This exact problem is also reported in \cite{discon} where it is suggested that PMP requires the state to be continuous. Caution is therefore required when tuning $\phi$. 

\subsection{Test case 2: the ETC driving cycle}
\begin{figure*}[htb]
    \centering
    \includegraphics[width=1\linewidth]{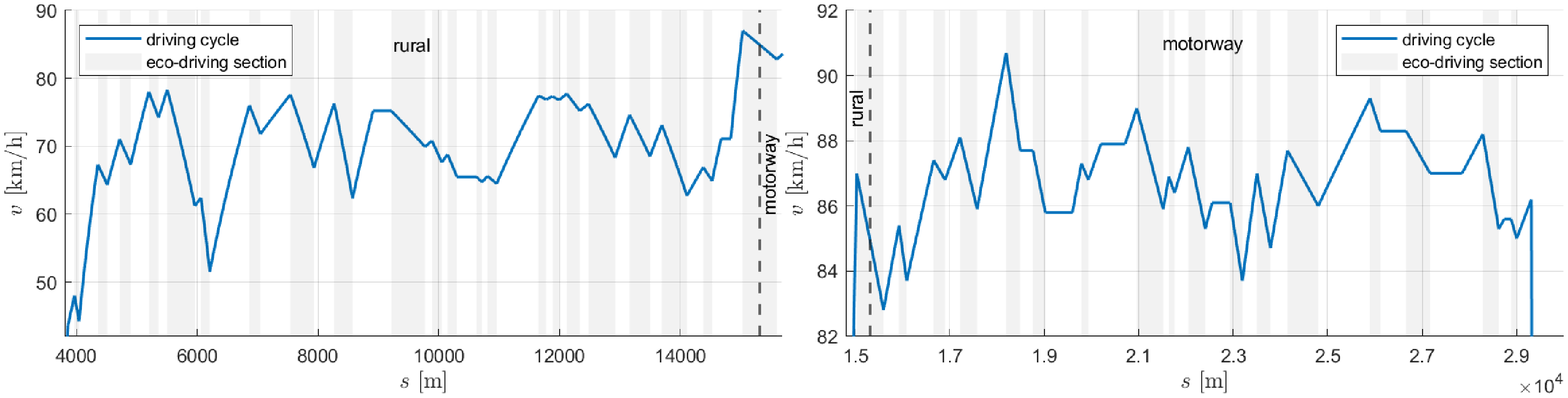} 
    \caption{Velocity profile of the rural part (left) and the motorway part (right) of the ETC driving cycle \cite{etcbook} with ED sections indicated in gray.}
    \label{fig:tc2fig}
\end{figure*}

Test case 1 showed that eco--driving has a high potential to achieve  significant fuel savings under driving scenarios that favor the use of ED and provided useful insights in the effect of changing the cost function weights.
To provide a more realistic estimate of the performance of the developed ED PMP algorithm, the European transient cycle driving cycle \cite{etcbook} is selected, which contains also sections of acceleration and cruising on an extended route. 
This reference driving cycle for ICE trucks consists of 3 parts: urban, rural and motorway, each 600~[s] in duration. The latter 2 parts are considered in this test case and depicted in Fig.~\ref{fig:tc2fig}. Around 45.7\% of the rural part length consist of sections where deceleration eco--driving advice can be given, whereas the motorway equivalent is only 36.9\%.
%
%
The ETC was divided into sections of acceleration, cruising and deceleration, after which it was piecewise simulated for optimal velocity reference tracking. 

Table~\ref{tab:tc2simresults25200} presents nominal and relative results for both ETC parts. One can conclude from Table~\ref{tab:tc2simresults25200} that relatively more fuel is saved with eco--driving advice in the rural part than in the motorway part, which is explained by the amount of eco--driving sections and higher velocity fluctuations. From test case 1 it was concluded that a lower $\phi$ results in more fuel savings. This is supported by data of the rural part, but not by the data of the motorway part of the ETC. The fuel savings are approximately constant with the highest amount of fuel saved at the highest $\phi$. Furthermore, the conclusions from test case 1 suggest that the higher average velocity of the motorway part should result in higher fuel savings for a constant $\phi$ to some extent. Instead, the opposite holds with the most fuel savings in the rural part. 
These findings suggest that a true assessment of the performance in the real world is more complex and is also, to some degree, proportional to the average velocity, the variability in velocity and the possibility to apply eco--driving advice. 
Note also that the ETC driving cycle is intended for ICE dynamometer testing, whereas in this paper it has been used for chassis dynamometer testing. In general, the obtained results indicate that EDAS will improve fuel savings proportionally to the number of ED sections and difference in initial and final velocity, and depending on the cost weights ratio. 
\setlength{\tabcolsep}{0.29em}
\begin{table}[htb]
    \caption{Test case 2: trip time increase, fuel consumption and saved fuel (time increase and saved fuel w.r.t. ED disabled).}
    \centering
    \begin{tabular}{ @{}cc|cccc|cccc@{\hspace{-0.05mm}}}
         \hline
        & & \multicolumn{4}{c|}{rural part} & \multicolumn{4}{c}{motorway part}\\
        \cline{3-10}
ED       & $\phi$ & time  & incr. & fuel   & saved & time  & incr. & fuel   & saved\\
        & {[}-]  & [s]   & [\%]  & [g]    & [\%]  & [s]   & [\%]  & [g]    & [\%]\\ \hline
    disabled&[-]&595.13&[-]&3346.17&[-]&577.34&[-]&3309.15&[-]\\
    enabled&15&601.43&1.06&2796.81&16.42 &577.53&0.04&3231.73&2.34\\
    enabled&30&601.01&0.99&2808.82&16.06 &577.52&0.03&3243.02&2.00\\
    enabled&60&598.27&0.53&2942.99&12.05 &577.58&0.04&3231.08&2.36\\
         \hline
    \end{tabular}
\label{tab:tc2simresults25200}
\end{table}

\section{Conclusions} \label{sec:confw}

In this paper, an eco--driving PMP algorithm for optimal deceleration and gear shifting in trucks has been designed and implemented in IPG TruckMaker \cite{TM}. The developed EDAS strategy computes an optimal sequence of driving modes for reducing fuel consumption when decelerating and incorporates contextual feedback from the driver for safety. Furthermore, the optimization over driving modes is computationally inexpensive, allowing the methodology to be used online, in real--time electronic control units. 
Performance has been assessed in two test cases that showed that the developed ED PMP algorithm leads to reduced fuel consumption with a small increase in trip duration. 
In the simulations, an ideal driver that applies the advice instantaneously has been considered. Human driver behavior, e.g., delayed reaction, can be further considered to obtain a more realistic performance assessment. Moreover, velocity--dependent cost weights can be used to enable more consistent fuel savings along routes with velocity profiles of high variability. 
It is of interest to further validate the developed PMP algorithm for eco--driving by using the VECTO tool.

\section*{Appendix}
\setlength{\tabcolsep}{0.2em}
\begin{table}[htb]
\caption{Utilized truck parameters provided in IPG TruckMaker~\cite{TM}.}
\centering
\begin{threeparttable}
  \begin{tabular}{|cc|cc|cc|}
  \toprule
  Parameter&Value&Parameter&Value&Parameter&Value\\
  \midrule
  $m$[kg] & 25200 &$i_t(y=0)$[-] &0&$J_{eq.dt}$[kgm\textsuperscript{2}]&83.77 \\
  $c_r$[-] & 9.57e-3 &  $i_t(y=1)$[-] &14.12& $J_{eq.pt}$[kgm\textsuperscript{2}]& 3983.74\\
  $c_d$[-]&0.41 & $i_t(y=2)$[-] &9.54& $J_{eq.pt}$[kgm\textsuperscript{2}]&1864.56 \\
  $A_f$[m\textsuperscript{2}]&10.2&$i_t(y=3)$[-] &6.52& $J_{eq.pt}$[kgm\textsuperscript{2}]&916.41 \\
  $i_r$[-]&3.08&$i_t(y=4)$[-] &4.75& $J_{eq.pt}$[kgm\textsuperscript{2}]&526.28 \\
  $r_w$[m] &0.496&  $i_t(y=5)$[-] &3.09& $J_{eq.pt}$[kgm\textsuperscript{2}]&271.06 \\
  $\omega_{idle}$[RPM]&450&  $i_t(y=6)$[-] &2.09& $J_{eq.pt}$[kgm\textsuperscript{2}]&169.60 \\
  $\dot{m}_{f,idle}$[gs\textsuperscript{-1}]&0.09542&  $i_t(y=7)$[-] &1.43& $J_{eq.pt}$[kgm\textsuperscript{2}]&123.97 \\
  $T_{eb}$[Nm]& $\approx$1278&  $i_t(y=8)$[-] &1& $J_{eq.pt}$[kgm\textsuperscript{2}]&103.42 \\
  $\beta_0$[-] &0.3615&$\beta_2$[-] &5.816e-4&$\beta_4$[-] &5.866e-6\\
  $\beta_1$[-] &  -8.521e-4& $\beta_3$[-] &4.489e-7& $\beta_5$[-] &-4.083e-7 \\
$\gamma_0$[-]&-16.87&$\gamma_1$[-]&0.2899& fuel & diesel\\
  \bottomrule
  \end{tabular}
  \label{tab:prop}
 \end{threeparttable}
\end{table}

\bibliographystyle{IEEEtran}
\bibliography{references}

\end{document}